\documentclass[aps,prb,twocolumn,floats]{revtex4}
\usepackage{graphics}
\usepackage{amssymb}
\usepackage{amsmath}
\usepackage{mathtools}

\begin{document}

\title{Quadrupolar interactions between acceptor pairs in p-doped semiconductors}
\author{Adam C. Durst$^1$, Genesis Yang-Mejia$^1$, and R. N. Bhatt$^{2,3}$}
\affiliation{$^1$Department of Physics and Astronomy, Hofstra University, Hempstead, NY 11549-1510}
\affiliation{$^2$Department of Electrical Engineering, Princeton University, Princeton, NJ 08544}
\affiliation{$^3$School of Natural Sciences, Institute for Advanced Study, Princeton, NJ 08540}
\date{October 15, 2019}

\begin{abstract}
We consider the interaction between acceptor pairs in doped semiconductors in the limit of large inter-acceptor separation relevant for low doping densities.  Modeling individual acceptors via the spherical model of Baldereschi and Lipari, we calculate matrix elements of the quadrupole tensor between the four degenerate ground states and show that the acceptor has a nonzero quadrupole moment.  As a result, the dominant contribution to the large-separation acceptor-acceptor interaction comes from direct (charge-density) terms rather than exchange terms.  The quadrupole is the leading nonzero moment, so the electric quadrupole-quadrupole interaction dominates for large separation.  We calculate the matrix elements of the quadrupole-quadrupole interaction Hamiltonian in a product-state basis and diagonalize, obtaining a closed-form expression for the energies and degeneracies of the sixteen-state energy spectrum.  All dependence on material parameters enters via an overall prefactor, resulting in surprisingly simple and universal results.  This simplicity is due, in part, to a mathematical happenstance, the nontrivial vanishing of a particular Wigner 6-$j$ symbol, $\left\{ \begin{array}{ccc} 2 & 2 & 2 \\ \tfrac{3}{2} & \tfrac{3}{2} & \tfrac{3}{2} \end{array} \right\} = 0$.  Results are relevant to the control of two-qubit interactions in quantum computing implementations based on acceptor spins, as well as calculations of the thermodynamic properties of insulating $p$-type semiconductors.
\end{abstract}

\maketitle

\section{Introduction}
\label{sec:intro}
A leading candidate for implementing qubits for quantum computation is the use of dopant spins in silicon and other semiconductors \cite{kan98,zwa13,tah02,mor10,tyr12,sae13,pic14a,pic14b,kaw14,kim15,pic16a,pic16b,mi17,gol03,rus13,van14,aba16,sal16a,sal16b,aba17,van17}.  In such implementations, controlling two-qubit interactions requires a detailed understanding of the dopant-dopant interaction and its dependence on the separation between dopants.  One of the difficulties associated with donor-based qubits in multivalley semiconductors like Si, Ge, and AlAs is that the donor-donor interaction has a significant oscillatory component as a function of inter-donor separation.  Since the oscillation occurs on atomic lengthscales, control of donor-donor interactions often requires precise placement of dopant atoms, which can be problematic \cite{pic14a}.  Acceptor-based qubits \cite{gol03,rus13,van14,aba16,sal16a,sal16b,aba17,van17} lack such multivalley complications, so variation in the acceptor-acceptor interaction occurs on the much longer lengthscale of the effective Bohr radius (tens to hundreds of angstroms).

However, while donors are well modeled as effective hydrogen atoms and donor pairs as effective hydrogen molecules, acceptors in tetrahedrally-coordinated semiconductors are somewhat more complex, due to a degenerate valence band maximum and the effects of spin-orbit coupling, as described by the Luttinger Hamiltonian \cite{lut56,koh57}.  Acceptor models that account for these effects were studied a long time ago \cite{sch62,men64,bal73,bal74}.  One particularly useful formulation is due to Baldereschi and Lipari \cite{bal73}, who showed that the acceptor problem based on the Luttinger Hamiltonian can be reformulated so as to split it into two parts.  These parts, which correspond to different behavior in angular momentum space, consist of a ``spherical'' term, which can be solved quite accurately, yielding much better acceptor ground states than earlier variational estimates, plus a ``cubic'' correction, which can be treated perturbatively \cite{bal74}.  In what follows, we refer to the model based on the first term alone, valid when cubic corrections can be neglected, as the Baldereschi-Lipari spherical model.  Note that this model is ``spherical'' in the sense that its Hamiltonian is spherically symmetric, like an atomic system.  However, this does not preclude the eigenfunctions from having nontrivial spatial angular structure, which they do.  In fact, the four-fold degenerate Baldereschi-Lipari ground-state acceptor wave functions consist of two terms, an $s$-wave term, reminiscent of the ground state of hydrogen, as well as a $d$-wave term, which exists due to strong spin-orbit coupling.

In prior work \cite{dur17}, we used the Baldereschi-Lipari single-acceptor wave functions to develop a Heitler-London model for the acceptor pair.  This numerical calculation provided the acceptor-pair energy spectrum for input values of material parameters and inter-acceptor separation, but due to computational constraints, its results were limited to acceptors separated by less than a few effective Bohr radii.

In the present work, we perform a complementary calculation of acceptor-pair energy spectra, valid in the large-separation limit, and yielding closed-form solutions.  Doing so was necessary, as many applications require an understanding of acceptor-pair interactions over a wide range of inter-acceptor separations.  In particular, this large-separation limit is directly relevant to quantum computing applications, where dopant concentrations are typically dilute.  It is also relevant to calculations of the thermodynamic properties of $p$-type semiconductors, which involve a system of many randomly-distributed acceptors, separated from each other by distances that vary by large factors.

The fact that the Baldereschi-Lipari single-acceptor wave functions contain both $s$-wave and $d$-wave terms means that the charge distribution of a ground-state acceptor has a nonzero quadrupole moment, in contrast to the spherically symmetric charge distribution of a hydrogenic ground-state donor.  As a result, while the dominant contribution to the large-separation donor-donor interaction comes from an exchange term and therefore decays exponentially with inter-donor separation, the dominant contribution to the large-separation acceptor-acceptor interaction comes from a direct (charge-density) term \cite{bha79} and therefore decays as a power law with inter-acceptor separation.  Since the first nonzero multipole moment of the acceptor is the quadrupole moment, we calculate, in this paper, the quadrupole-quadrupole interaction between well-separated acceptors.

We formulate our model in Sec.~\ref{sec:model}, calculate matrix elements of the quadrupole tensor in Sec.~\ref{sec:quadtensor}, and use those matrix elements to calculate the quadrupole-quadrupole energy spectrum in Sec.~\ref{sec:quadquad}.  Details of our matrix element calculations are presented in the Appendices.  Due, in part, to a mathematical happenstance, the nontrivial vanishing of a particular Wigner 6-$j$ symbol, our results turn out to be far simpler than would generically be expected.  Aside from the variational coefficients of the Baldereschi-Lipari wave functions, which are computed numerically, all other calculations are performed analytically with results presented in closed form.  Conclusions are discussed in Sec.~\ref{sec:conclusions}.

\section{Model}
\label{sec:model}

\subsection{Single Acceptor Model}
\label{ssec:baldlip}
We model each acceptor via the spherical model developed by Baldereschi and Lipari \cite{bal73}, which treats the acceptor-ion-plus-hole system as a hydrogenic atom modified to account for valence band degeneracy and spin-orbit coupling.  In tetrahedrally-coordinated semiconductors, the low-energy band structure consists of a nondegenerate (aside from spin) conduction band minimum and a degenerate valence band maximum.  Spin-orbit coupling breaks a three-fold degeneracy down to two-fold, with a split-off bottom band that can be safely neglected in the large coupling limit, which we will assume.  (Large coupling is a good approximation for most semiconductors, though less so for Si.)  Including spin, this leaves a four-fold degeneracy at the top of the valence band, compared to the two-fold spin degeneracy at the bottom of the conduction band.  Baldereschi and Lipari model the four-fold degenerate holes via an effective spin $J=3/2$, with $J_z=\{-3/2,-1/2,1/2,3/2\}$ labeling the four degenerate states.  Thus, the acceptor problem becomes that of a spin-$3/2$ particle in the presence of a Coulomb potential and spin-orbit coupling.  Within the effective mass approximation, the Hamiltonian can be written as \cite{lut56,koh57,bal73}
\begin{eqnarray}
H &=& (\gamma_{1} + \frac{5}{2} \gamma_{2}) \frac{p^{2}}{2m_{0}}
- \frac{\gamma_{2}}{m_{0}} \left( p_{x}^{2} J_{x}^{2} + p_{y}^{2} J_{y}^{2}
+ p_{z}^{2} J_{z}^{2} \right) - \frac{e^{2}}{\epsilon_{0} r} \nonumber \\
&& - \frac{2\gamma_{3}}{m_{0}} \Big( \{p_{x},p_{y}\} \{J_{x},J_{y}\}
+ \{p_{y},p_{z}\} \{J_{y},J_{z}\} \nonumber \\
&& \;\;\;\;\;\;\;\;\;\;\;\;\;\;\;\;\;\;\;\;\;\;\;\;\;\;\;\;\;\;\;\;\;\;\;\;\;
+ \{p_{z},p_{x}\} \{J_{z},J_{x}\} \Big)
\label{eq:KLhamiltonian}
\end{eqnarray}
which is known as the Luttinger Hamiltonian \cite{lut56}.  Here $\{a,b\} \equiv (ab + ba)/2$, ${\bf J}$ is the hole angular momentum operator corresponding to spin-3/2, ${\bf p}$ is the hole momentum operator, $m_{0}$ is the free electron mass, $\epsilon_{0}$ is the crystal dielectric constant, and $\gamma_{1}$, $\gamma_{2}$, and $\gamma_{3}$ are the Luttinger constants describing hole dispersion near the top of the valence band.  This expression has the cubic symmetry of the semiconductor crystal.  The innovation of Baldereschi and Lipari \cite{bal73} was to rewrite it in such a way that the terms with full spherical symmetry are separated from those with only cubic symmetry.  Expressing energies in units of the effective Rydberg, ${\rm Ryd} \equiv e^4 m_0/2\hbar^2\epsilon_0^2\gamma_1$, and lengths in units of the effective Bohr radius, $a_B \equiv \hbar^2\epsilon_0\gamma_1/e^2m_0$, they showed that
\begin{eqnarray}
H &=& - \nabla^{2} - \frac{2}{r} - \frac{\mu}{9 \hbar^{2}}
\left( P^{(2)} \cdot J^{(2)} \right) \nonumber \\
&& + \frac{\delta}{9 \hbar^{2}}
\Big( \left[ P^{(2)} \times J^{(2)} \right]^{(4)}_{4}
+ \frac{\sqrt{70}}{5} \left[ P^{(2)} \times J^{(2)} \right]^{(4)}_{0}
\nonumber \\
&& \;\;\;\;\;\;\;\;\;\;\;\;\;\;\;\;\;\;\;\;\;\;\;\;\;\;\;\;\;\;\;\;\;\;\;\;\;
+ \left[ P^{(2)} \times J^{(2)} \right]^{(4)}_{-4} \Big)
\label{eq:cubichamiltonian}
\end{eqnarray}
where $P^{(2)}$ and $J^{(2)}$ are rank-2 spherical tensor operators for momentum and effective spin (see Refs.~\onlinecite{bal73,edm57} for details regarding the irreducible tensor notation).  The third term, proportional to $\mu \equiv (6 \gamma_{3} + 4 \gamma_{2}) / 5 \gamma_{1}$, is the spherical contribution to the spin-orbit interaction, and the fourth term, proportional to $\delta \equiv (\gamma_{3} - \gamma_{2}) / \gamma_{1}$, is the cubic contribution.  In nearly all semiconductors (with the important exception of Si), $\delta$ is much smaller than $\mu$, so the cubic term can be safely neglected.  Doing so yields the Baldereschi-Lipari \cite{bal73} spherical Hamiltonian
\begin{equation}
H = -\nabla^2 - \frac{2}{r} - \frac{\mu}{9\hbar^{2}} \left( P^{(2)} \cdot J^{(2)}\right)
\label{eq:hamiltonian}
\end{equation}
which depends on material parameters only through its units, the effective Rydberg and Bohr radius, as well as $\mu$, a dimensionless parameter between 0 and 1 that indicates the strength of spin-orbit coupling in the material.

Since this Hamiltonian is spherically symmetric by construction, total angular momentum, ${\bf F} = {\bf L} + {\bf J}$, is conserved, as in an atomic system, where ${\bf L}$ is the orbital angular momentum and ${\bf J}$ is the effective spin angular momentum of the hole.  Since the spin-orbit term couples states of $\Delta L = 0,\pm 2$, the most general expression for the acceptor ground-state wave function is: \cite{bal73}
\begin{eqnarray}
|\Psi_{F_z} \rangle
&=& f_0(r) \left| L=0, J=\tfrac{3}{2}, F=\tfrac{3}{2},F_z \right\rangle \nonumber \\
&+& g_0(r) \left| L=2,J=\tfrac{3}{2}, F=\tfrac{3}{2},F_z \right\rangle
\label{eq:groundstate}
\end{eqnarray}
where the $|LJFF_z\rangle$ kets are eigenfunctions of total angular momentum and $f_0(r)$ and $g_0(r)$ are radial functions.

It is important to note that while the Baldereschi-Lipari Hamiltonian [Eq.~(\ref{eq:hamiltonian})] is spherically symmetric, individual eigenfunctions of the Hamiltonian, and their associated charge density distributions, are not necessarily spherically symmetric (as is the case for the hydrogen atom, where all eigenfunctions except the $s$-orbitals lack the spherical symmetry of the hydrogenic Hamiltonian).  A key difference between the hydrogen atom and the Baldereschi-Lipari acceptor, however, is in the symmetry of their ground states.  While the two degenerate ground states of the hydrogen atom are indeed spherically symmetric, the four degenerate ground states of the Baldereschi-Lipari acceptor are not spherically symmetric, due to the $L$=2 term in Eq.~(\ref{eq:groundstate}).

\begin{figure}
\centerline{\resizebox{3.5in}{!}{\includegraphics{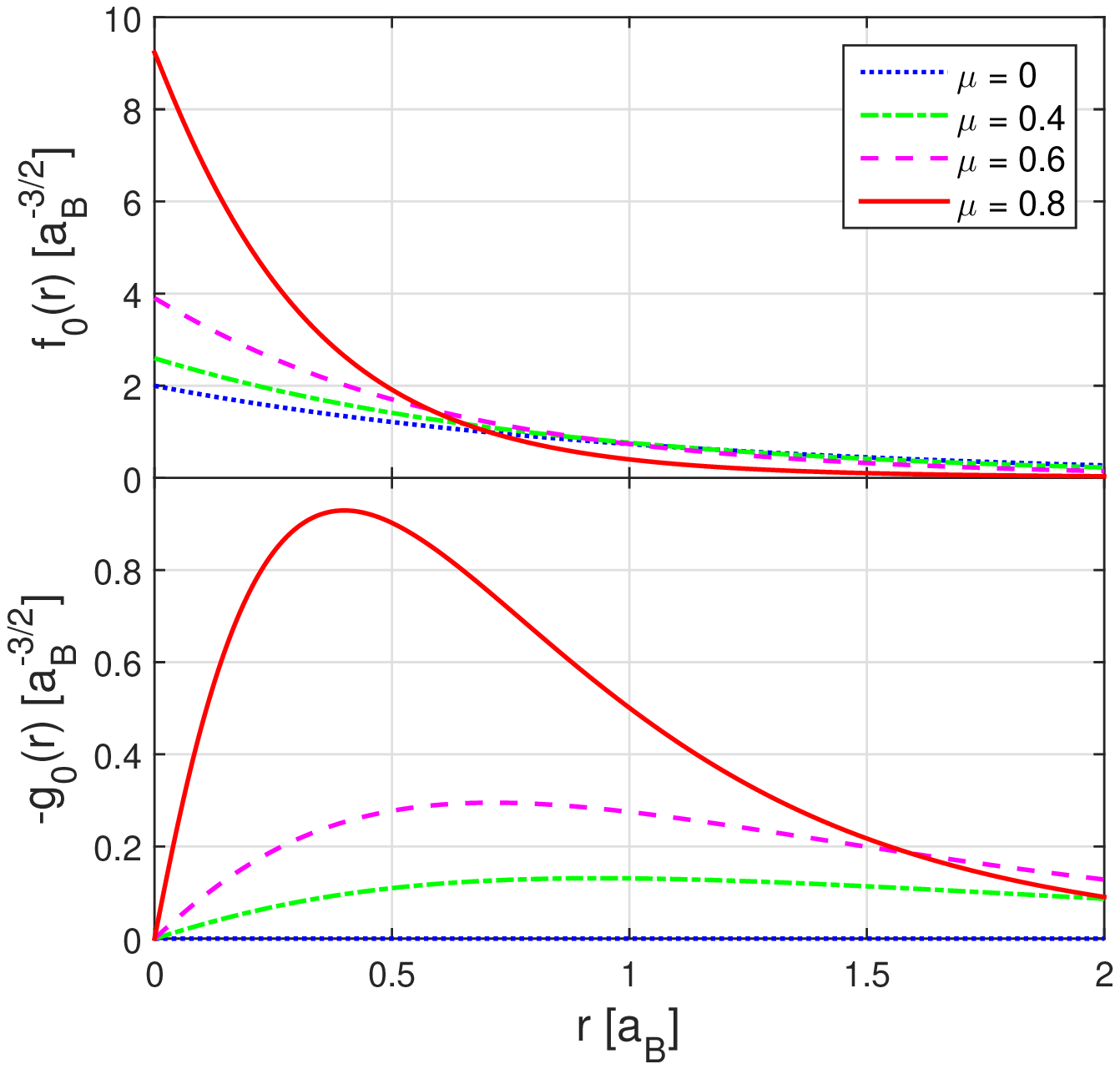}}}
\caption{Baldereschi-Lipari ground-state acceptor radial functions, $f_0$ and $g_0$, as a function of distance $r$ from the acceptor center, for various values of the spin-orbit coupling parameter $\mu$.  With increasing $\mu$, $g_0$ grows and both $f_0$ and $g_0$ become more localized.  These numerical results reproduce those of Ref.~\onlinecite{bal73}, Fig.~4.}
\label{fig:f0g0vsr}
\end{figure}

Following Baldereschi and Lipari, we use a variational approach to estimate the radial functions, employing trial radial functions of the form
\begin{equation}
f_0(r) = \sum\limits_{i = 1}^{21} A_i e^{-\alpha_i r^2} \;\;\;\;\;\;
g_0(r) = r\sum\limits_{i=1}^{21} B_i e^{-\alpha_i r^2}
\label{eq:radialfunctions}
\end{equation}
where the constants $\alpha_i$ are chosen in geometric progression $(\alpha_{i+1} = g\alpha_i)$ from $\alpha_1 = 10^{-2}$ to $\alpha_{21} = 5 \times 10^5$.  We numerically compute the ground-state energy, as well as the 42 variational parameters $A_i$ and $B_i$, as a function of spin-orbit coupling parameter $\mu$, by minimizing the expectation value of the Hamiltonian with respect to all 42 parameters.  The details of this procedure are provided in Appendix B of Ref.~\onlinecite{dur17}.  The resulting radial functions are plotted in Fig.~\ref{fig:f0g0vsr} and reproduce the results of Ref.~\onlinecite{bal73}.  For $\mu=0$, $f_0$ is the ground-state radial wave function of hydrogen and $g_0$ is zero.  As $\mu$ increases toward one, $g_0$ grows and both $f_0$ and $g_0$ become more localized.

The ground-state energy and radial functions are independent of the magnetic quantum number $F_z$.  Thus, Eq.~(\ref{eq:groundstate}) represents four degenerate ground states, labeled by $F_z=\{-3/2,-1/2,1/2,3/2\}$.  In what follows, we restrict our Hilbert space to this ground-state manifold and use these four Baldereschi-Lipari ground-state wave functions as a basis for evaluating matrix elements at low energies.

\subsection{Acceptor Interaction Model}
\label{ssec:intmodel}
To model the interaction between two acceptors, we construct acceptor-pair basis functions from the Baldereschi-Lipari ground-state acceptor wave functions of Eq.~(\ref{eq:groundstate}).  Since the acceptor-pair basis functions must be antisymmetric upon exchange of identical holes, they take the form
\begin{equation}
|F_{z1} F_{z2}\rangle = \tfrac{1}{\sqrt 2} \left( |\Psi_{F_{z1}}^A\rangle |\Psi_{F_{z2}}^B\rangle - |\Psi_{F_{z1}}^B\rangle |\Psi_{F_{z2}}^A\rangle \right)
\label{eq:antisymmetricwavefunction}
\end{equation}
where $|\Psi_{F_{z1}}\rangle$ and $|\Psi_{F_{z2}}\rangle$ are the single-acceptor states for acceptor 1 and acceptor 2 respectively, and the superscripts indicate which hole ($A$ or $B$) is in which state.  When evaluating matrix elements in this two-term basis, we obtain two direct (charge-density) terms where each state is occupied by the same hole in both the bra and the ket, and two exchange (cross) terms where the states are occupied by different holes in the bra versus the ket.  Since the single-acceptor states are localized to their acceptor locations on the scale of the effective Bohr radius, $a_B$, exchange terms decay exponentially with inter-acceptor separation, $R$, and can be neglected in comparison to direct terms (which decay as a power law) in the large-separation ($R \gg a_B$) limit that we consider herein.  Neglecting exchange terms is equivalent to treating the holes as distinguishable, which they effectively become for $R \gg a_B$.  Thus, the acceptor-pair basis functions reduce to simple product states
\begin{equation}
|F_{z1} F_{z2}\rangle \approx |\Psi_{F_{z1}}\rangle |\Psi_{F_{z2}}\rangle
\label{eq:productstates}
\end{equation}
where one hole is localized about acceptor 1 and the other is localized about acceptor 2.

In this product state basis, the Coulomb interaction between the two acceptors can be expressed in terms of a multipole expansion \cite{jac75} of the acceptor charge distribution.  The monopole moment vanishes because acceptors are neutral (total hole charge cancels the charge of the acceptor ion).  The dipole moment vanishes because electric dipole selection rules \cite{gri05} require bra and ket to differ by one in azimuthal quantum number, $\Delta L = \pm 1$, while the Baldereschi-Lipari ground-state wave functions [Eq.~(\ref{eq:groundstate})] consist of only $L$=0 and $L$=2 terms.  Since electric quadrupole selection rules allow $\Delta L = \{0,\pm 2\}$, we expect a nonzero quadrupole moment.  The leading term in the multipole expansion of our interaction Hamiltonian is therefore the quadrupole-quadrupole term.  Hence, in the large-separation limit, we can model the interaction between two acceptors by their electric quadrupole-quadrupole interaction.

\section{Matrix Elements of the Quadrupole Tensor}
\label{sec:quadtensor}

\subsection{Quadrupole Tensor}
\label{ssec:tensor}
The quadrupole tensor, $\tensor{\bf Q}$, is a traceless, symmetric, Cartesian tensor of rank 2.  In Cartesian coordinates, measured from the center of the acceptor, it takes the form \cite{jac75}
\begin{equation}
\tensor{\bf Q}=
\left[
\begin{array}{ccc}
3x^2 - r^2      & 3xy 				& 3xz \\
3yx  			& 3y^2 - r^2        & 3yz \\
3zx 			& 3zy 	            & 3z^2 - r^2
\end{array}
\right] .
\label{eq:cartesiantensor}
\end{equation}
Rewriting the above in spherical coordinates, each component can be expressed as a linear combination of spherical harmonics \cite{gri05}, $Y_\ell^m(\theta,\phi)$, with $\ell=2$.
\begin{eqnarray}
\tensor{\bf Q} &=& \sqrt{\frac{2\pi}{5}} r^2 \left\{
\sqrt{2} Y_2^0 \left[ \begin{array}{ccc} -1 & 0 & 0 \\ 0 & -1 & 0 \\ 0 & 0 & 2 \end{array} \right] \right. \nonumber \\
&+& \!\!\! \left. \sqrt{3} Y_2^1 \left[ \begin{array}{ccc} 0 & 0 & -1 \\ 0 & 0 & i \\ -1 & i & 0 \end{array} \right]
+ \sqrt{3} Y_2^{-1} \left[ \begin{array}{ccc} 0 & 0 & 1 \\ 0 & 0 & i \\ 1 & i & 0 \end{array} \right] \right. \nonumber \\
&+& \!\!\! \left. \sqrt{3} Y_2^2 \left[ \begin{array}{ccc} 1 & -i & 0 \\ -i & -1 & 0 \\ 0 & 0 & 0 \end{array} \right]
+ \sqrt{3} Y_2^{-2} \left[ \begin{array}{ccc} 1 & i & 0 \\ i & -1 & 0 \\ 0 & 0 & 0 \end{array} \right] \right\}
\label{eq:sphericalharmonicstensor}
\end{eqnarray}
Since the Baldereschi-Lipari ground-state acceptor wave functions [Eq.~(\ref{eq:groundstate})] are linear combinations of coupled angular momentum eigenstates, $|L J F F_z\rangle$, and the components of the quadrupole tensor are linear combinations of $\ell$=2 spherical harmonics, $Y_2^m$, the evaluation of the matrix elements of the quadrupole tensor between Baldereschi-Lipari wave functions reduces to the evaluation of the matrix elements of the $\ell$=2 spherical harmonics between such angular momentum eigenstates.

\subsection{Matrix Elements}
\label{ssec:matrixelements}
\begin{figure}
\centerline{\resizebox{3.5in}{!}{\includegraphics{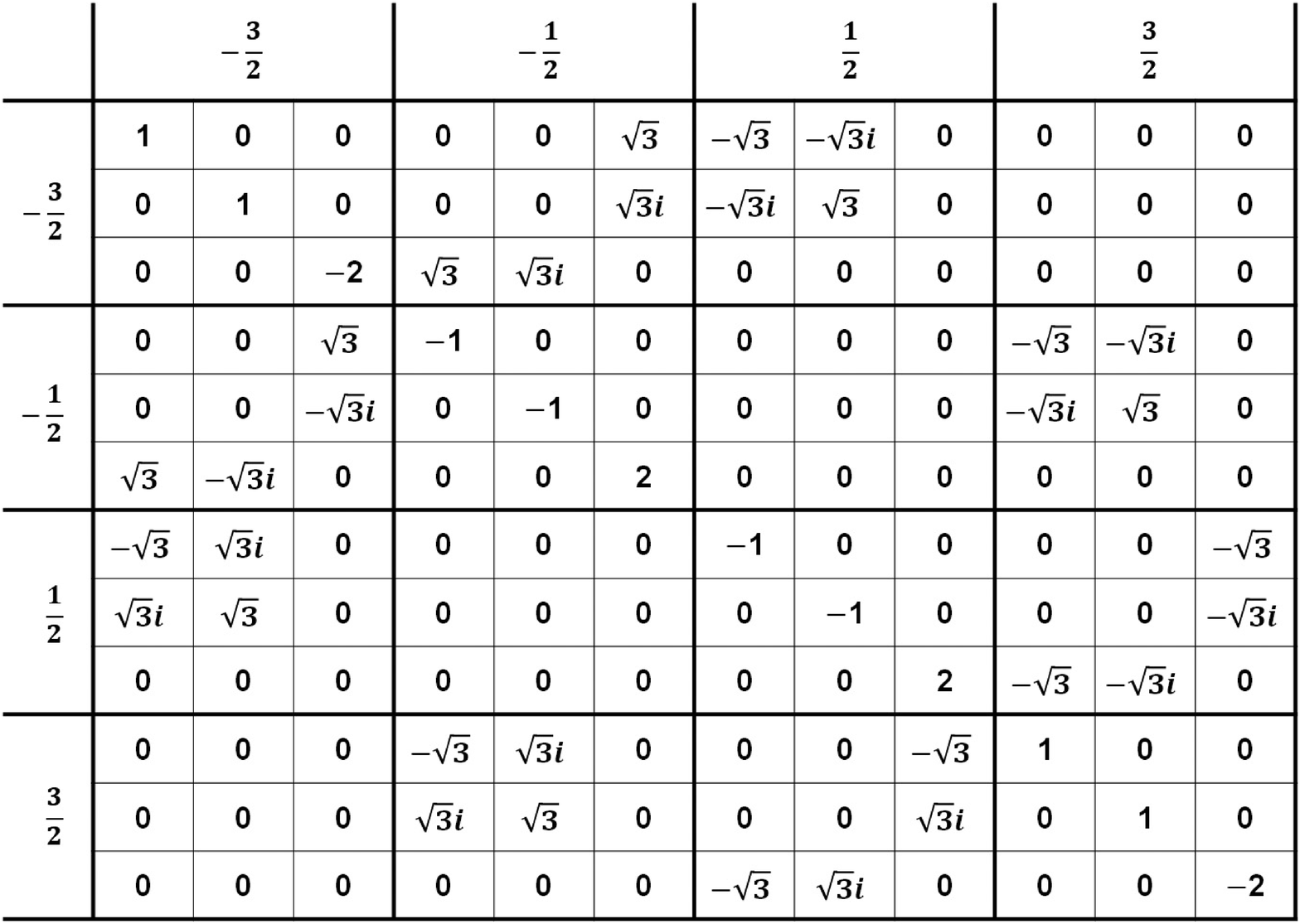}}}
\caption{Matrix elements of the quadruple tensor between Baldereschi-Lipari ground-state acceptor wave functions, $\langle \Psi_{F'_z} | \tensor{\bf Q} | \Psi_{F_z} \rangle$, are obtained by multiplying this $4 \times 4$ matrix of $3 \times 3$ tensors by a common factor of $Q_0/2 = -\tfrac{2}{5} R_{fg}$.  $F'_z$ and $F_z$ label the rows and columns of the matrix respectively.}
\label{fig:tensorsmatrix}
\end{figure}

The matrix elements of $r^2 Y_2^m(\theta,\phi)$ between the Baldereschi-Lipari ground-state acceptor wave functions [Eq.~(\ref{eq:groundstate})] take the form
\begin{eqnarray}
\langle \Psi_{F'_z} | r^2 Y_2^m | \Psi_{F_z} \rangle &=& \langle 0 \tfrac{3}{2} \tfrac{3}{2} F'_z | Y_2^m | 0 \tfrac{3}{2} \tfrac{3}{2} F_z \rangle R_{ff} \nonumber \\
&+& \langle 2 \tfrac{3}{2} \tfrac{3}{2} F'_z | Y_{2}^m | 0 \tfrac{3}{2} \tfrac{3}{2} F_z \rangle R_{fg} \nonumber \\
&+& \langle 0 \tfrac{3}{2} \tfrac{3}{2} F'_z | Y_{2}^m | 2 \tfrac{3}{2} \tfrac{3}{2} F_z \rangle R_{fg} \nonumber \\
&+& \langle 2 \tfrac{3}{2} \tfrac{3}{2} F'_z | Y_{2}^m | 2 \tfrac{3}{2} \tfrac{3}{2} F_z \rangle R_{gg}
\label{eq:matrixelement}
\end{eqnarray}
where $|0 \tfrac{3}{2} \tfrac{3}{2} F_z\rangle$ is shorthand for $|L=0, J=\tfrac{3}{2}, F=\tfrac{3}{2}, F_z\rangle$ and
\begin{eqnarray}
R_{ff} &\equiv& \int_0^{\infty} r^4 [f_0(r)]^2 dr \nonumber \\
R_{fg} &\equiv& \int_0^{\infty} r^4 f_0(r) g_0(r) dr \nonumber \\
R_{gg} &\equiv& \int_0^{\infty} r^4 [g_0(r)]^2 dr .
\label{eq:radialintegrals}
\end{eqnarray}
are radial integrals that depend on material parameters via the dependence of $f_0$ and $g_0$ on the spin-orbit coupling parameter $\mu$.  For simplicity, we refer to the four coupled-state matrix elements of $\ell$=2 spherical harmonics that appear in Eq.~(\ref{eq:matrixelement}) as the 0-0, 2-0, 0-2, and 2-2 terms, respectively.  We calculate these in Appendix~\ref{app:matrixelementcalculation} by making use of the Wigner-Eckart theorem (see Appendix~\ref{app:wigner}) and show that the 0-0 term is trivially zero, the 2-2 term is nontrivially zero (more about this in Sec.~\ref{ssec:vanish}), and the 2-0 and 0-2 terms are equal and nonzero.  Thus, Eq.~(\ref{eq:matrixelement}) simplifies to
\begin{equation}
\langle \Psi_{F'_z} | r^2 Y_2^m | \Psi_{F_z} \rangle = 2 \; \langle 2 \tfrac{3}{2} \tfrac{3}{2} F'_z | Y_2^m | 0 \tfrac{3}{2} \tfrac{3}{2} F_z \rangle \; R_{fg} .
\end{equation}
This result, combined with Eq.~(\ref{eq:matrices20term}), allows us to calculate the matrix elements of the five terms of the quadrupole tensor in Eq.~(\ref{eq:sphericalharmonicstensor}) between all combinations of the four degenerate ground states of the acceptor.  Doing so, we obtain a $4 \times 4$ matrix (row $F'_z$ versus column $F_z$) of $3 \times 3$ quadrupole tensors, given by an overall factor of $-\tfrac{2}{5} R_{fg}$ multiplying the matrix of tensors shown in Fig.~\ref{fig:tensorsmatrix}.
\begin{equation}
\langle \Psi_{F'_z} | \tensor{\bf Q} | \Psi_{F_z} \rangle = -\tfrac{2}{5} R_{fg} \times [{\rm Fig.~\ref{fig:tensorsmatrix}}]
\label{eq:quadrupolematrixelements}
\end{equation}
These quadrupole-tensor matrix elements will be used in Sec.~\ref{sec:quadquad} to calculate the quadrupole-quadrupole interaction between Baldereschi-Lipari acceptors.

\begin{figure}
\centerline{\resizebox{3.5in}{!}{\includegraphics{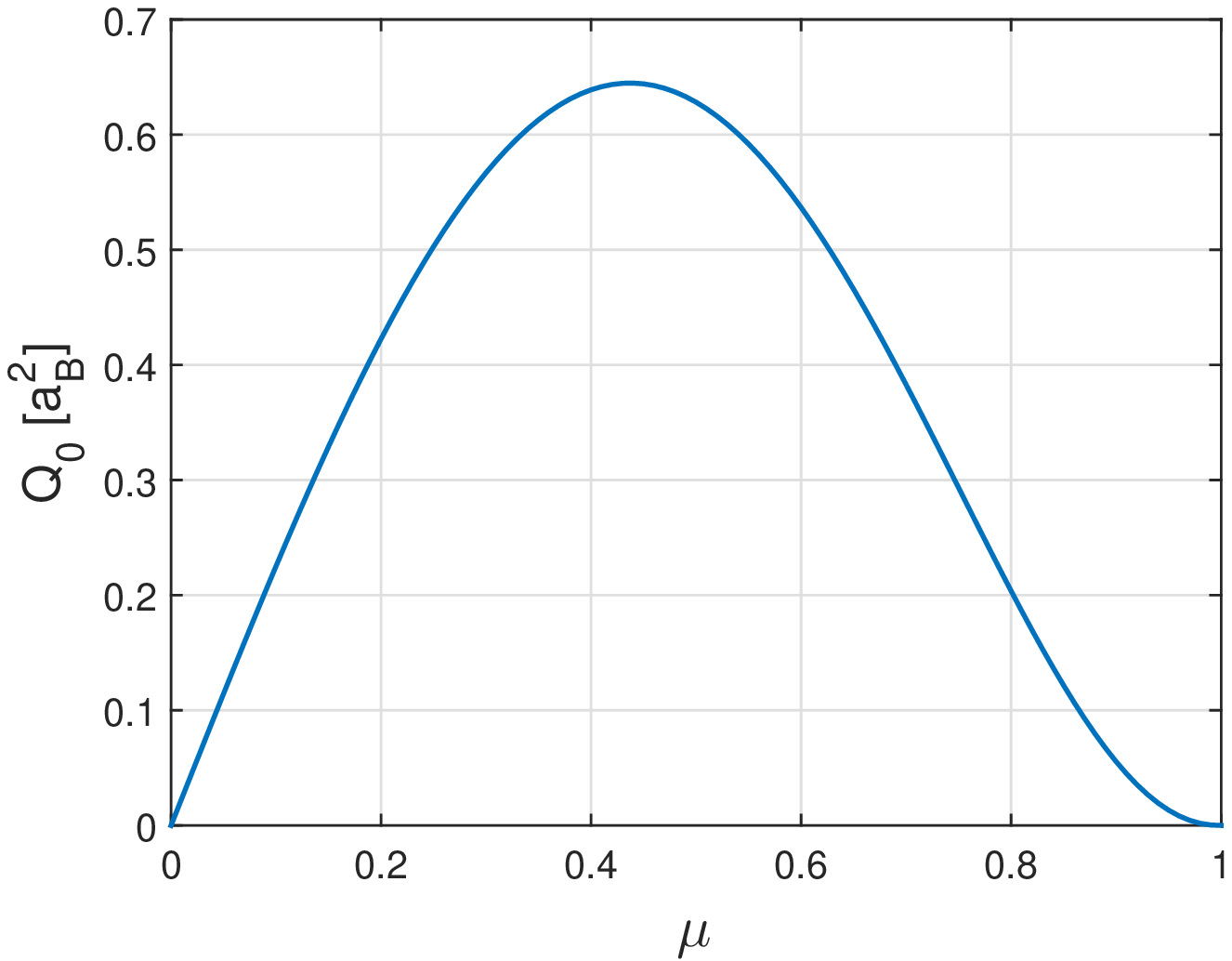}}}
\caption{Acceptor quadrupole moment, $Q_0 = -\tfrac{4}{5} R_{fg} = | \langle \Psi_{F_z} | Q_{zz} | \Psi_{F_z} \rangle |$, as a function of spin-orbit coupling parameter $\mu$.  Note that $Q_0$ is maximized for intermediate $\mu$, vanishing as $\mu \rightarrow 0$, since the wave function becomes pure $s$-wave, and as $\mu \rightarrow 1$, since the wave function becomes localized to the origin.}
\label{fig:Q0vsMu}
\end{figure}

It is convenient to name the $\mu$-dependent prefactor $Q_0/2$ and thereby define
\begin{eqnarray}
Q_0(\mu) &\equiv& -\tfrac{4}{5} R_{fg} = -\tfrac{4}{5} \int_0^{\infty} r^4 f_0(r) g_0(r) dr \nonumber \\
&=& -\tfrac{4}{5} \sum_{ij} \frac{A_i B_j}{(\alpha_i + \alpha_j)^3} \nonumber \\
&=& \left| \langle \Psi_{F_z} | Q_{zz} | \Psi_{F_z} \rangle \right|
\label{eq:Q0}
\end{eqnarray}
where the third equality results from expanding the radial functions via Eq.~(\ref{eq:radialfunctions}) and evaluating the Gaussian integrals, and the fourth equality is obtained by inspection of the diagonal $Q_{zz}$ components in Fig.~\ref{fig:tensorsmatrix}.  Since $Q_0$ is equal to the absolute value of the expectation value of $Q_{zz}$ in any of the four acceptor ground states, we hereafter refer to it as the quadrupole moment of the acceptor.  In Fig.~\ref{fig:Q0vsMu}, we plot $Q_0$ as a function of $\mu$.  Note that the acceptor quadrupole moment vanishes at both $\mu=0$ and $\mu=1$.  At $\mu=0$, it vanishes trivially because $g_0 = 0$. Without the $d$-wave ($L$=2) term in Eq.~(\ref{eq:groundstate}), the Baldereschi-Lipari wave function is pure $s$-wave and therefore lacks a quadrupole moment.  As $\mu \rightarrow 1$, $f_0$ and $g_0$ become sharply peaked about $r=0$ (see Fig.~\ref{fig:f0g0vsr}), so the $r^4$ factor in the integrand drives the integral to zero.  Essentially, the quadrupole moment vanishes as the angular structure of the wave function is squeezed to the origin.  Thus, the quadrupole moment of Baldereschi-Lipari acceptors is maximized when $\mu$ is large enough that the wave function has a significant $d$-wave component but $\mu$ is not so large that the wave function becomes too localized.

\subsection{Vanishing 2-2 Term}
\label{ssec:vanish}
As shown above (and in Appendix~\ref{app:matrixelementcalculation}), the matrix elements of the quadrupole tensor between Baldereschi-Lipari acceptor ground states have a particularly simple form.  Note that all dependence on material parameters (the Luttinger constants and the spin-orbit parameter $\mu$ derived therefrom) enter only via our units of length and energy (the effective Bohr radius and effective Rydberg) and an overall multiplicative factor, $R_{fg}$.  This simplicity is due, in part, to the vanishing of the $\langle 0 \tfrac{3}{2} \tfrac{3}{2} F_z^\prime | Y_2^m | 0 \tfrac{3}{2} \tfrac{3}{2} F_z \rangle$ and $\langle 2\tfrac{3}{2} \tfrac{3}{2} F_z^\prime | Y_2^m | 2 \tfrac{3}{2} \tfrac{3}{2} F_z \rangle$ terms in Eq.~(\ref{eq:matrixelement}), which multiply $R_{ff}$ and $R_{gg}$, respectively.  If these 0-0 and 2-2 terms did not vanish, then material parameters would enter via $R_{ff}$ and $R_{gg}$, in addition to $R_{fg}$.

There are a number of trivial ways that such $Y_\ell^m$ matrix elements between $|LJFF_z\rangle$ coupled states can vanish: (1) violation of the orbital-state triangle rule $|\ell-L|<L^\prime<\ell+L$, (2) orbital-state parity violation, which occurs when $\ell+L+L^\prime$ is odd, and (3) violation of the coupled-state triangle rule $|\ell-F|<F^\prime<\ell+F$.  In atomic physics, these define the selection rules that govern transitions between atomic states.  Viewed through the lens of the Wigner-Eckart theorem (see Appendix~\ref{app:wigner}), it is clear that all three of these cases result from the vanishing of Clebsch-Gordan coefficients, the ones in Eq.~(\ref{eq:orbitalwignereckart}), Eq.~(\ref{eq:orbitalreducedmatrixelements}), and Eq.~(\ref{eq:coupledwignereckart}), respectively.

The vanishing of our 0-0 term is, in this sense, trivial, due to a trivial zero of type (1), as per the classification in the prior paragraph.  However, the vanishing of our 2-2 term is not due to any of the trivial zero types enumerated above.  It is not due to the vanishing of a Clebsch-Gordan coefficient at all.  Rather, in this case, it is the coupled-state reduced matrix element from Eq.~(\ref{eq:coupledwignereckart}) that is itself zero.  As shown in Eqs.~(\ref{eq:22term}) through (\ref{eq:22zero}), it vanishes because up to three nonzero terms, each the product of three Clebsch-Gordan coefficients, happen to sum to zero.

To better understand this happenstance, it is helpful to notice that the sum of triple-products of Clebsch-Gordan coefficients that appears in Eq.~(\ref{eq:22term}) has precisely the form indicated in Eq.~(6.2.7) of Ref.~\onlinecite{edm57} and is therefore proportional to the product of a single Clebsch-Gordan coefficient and a 6-$j$ symbol (Racah coefficient).  Edmonds \cite{edm57} shows this to be true by making use of the definition of the 6-$j$ symbol and the orthogonality properties of the Clebsch-Gordan coefficients.  For the case at hand, we see that Eq.~(\ref{eq:22term}) becomes
\begin{eqnarray}
\lefteqn{\langle 2\tfrac{3}{2} \tfrac{3}{2} F_z^\prime | Y_2^m | 2 \tfrac{3}{2} \tfrac{3}{2} F_z \rangle} \nonumber \\
&& = -2\sqrt{5} C_{22}^2 \langle m F_z | \tfrac{3}{2} F_z^\prime \rangle_{2 \tfrac{3}{2}}
\left\{ \begin{array}{ccc} 2 & 2 & 2 \\ \tfrac{3}{2} & \tfrac{3}{2} & \tfrac{3}{2} \end{array} \right\}
\label{eq:22term6jsymbol}
\end{eqnarray}
where notational details are defined in Appendix~\ref{app:wigner} and the last factor is a 6-$j$ symbol.  That such matrix elements are proportional to a single 6-$j$ symbol turns out to be a general property of the matrix elements of spherical tensor operators between coupled angular momentum states where the spherical tensor commutes with one of the coupled angular momentum operators (here, the spin one) but not the other (here, the orbital one).  This is shown by Biedenharn and Louck (see Eq.~(3.246) of Ref.~\onlinecite{bie84a}).

The happenstance that results in the vanishing of our 2-2 term is that the particular 6-$j$ symbol that appears in Eq.~(\ref{eq:22term6jsymbol}) is equal to zero.
\begin{equation}
\left\{ \begin{array}{ccc} 2 & 2 & 2 \\ \tfrac{3}{2} & \tfrac{3}{2} & \tfrac{3}{2} \end{array} \right\} = 0
\label{eq:6jzero}
\end{equation}
The 6-$j$ symbols and the zeros thereof have been studied extensively in the mathematical physics literature \cite{bie84b,reg59,bar62,she64,koo74,rao84,bru85a,bru85b,bru85c,bre86,bru87,rao88,ray94}.  Their values are provided by the following general expression \cite{rao88,reg59,bie84a}:
\begin{equation}
\left\{ \begin{array}{ccc} j_1 & j_2 & j_3 \\ \ell_1 & \ell_2 & \ell_3 \end{array} \right\} = N \sum_{P=\alpha_{\rm max}}^{\beta_{\rm min}}
\frac{(-1)^P (P+1)!}{\displaystyle \prod_{i=1}^4 (P-\alpha_i)! \prod_{k=1}^3 (\beta_k-P)!}
\label{eq:6jgeneral}
\end{equation}
where $\alpha_{\rm max}$ is the largest of the four $\alpha_i$, $\beta_{\rm min}$ is the smallest of the three $\beta_i$,
\begin{eqnarray}
\alpha_1 = j_1 + j_2 + j_3 &\;\;& \beta_1 = j_1 + \ell_1 + j_2 + \ell_2 \nonumber \\
\alpha_2 = \ell_1 + \ell_2 + j_3 &\;\;& \beta_2 = j_1 + \ell_1 + j_3 + \ell_3 \nonumber \\
\alpha_3 = j_1 + \ell_2 + \ell_3 &\;\;& \beta_3 = j_2 + \ell_2 + j_3 + \ell_3 \nonumber \\
\alpha_4 = \ell_1 + j_2 + \ell_3 &\;\;&
\label{eq:6jalphabeta}
\end{eqnarray}
and
\begin{equation}
N = \Delta(j_1 j_2 j_3) \Delta(\ell_1 \ell_2 j_3) \Delta(j_1 \ell_2 \ell_3) \Delta(\ell_1 j_2 \ell_3)
\label{eq:6jN}
\end{equation}
\begin{equation}
\Delta(pqr) = \sqrt{\frac{(p+q-r)! (p-q+r)! (-p+q+r)!}{(p+q+r+1)!}}
\label{eq:6jDelta}
\end{equation}
where $\Delta(pqr)$ is zero if $p$, $q$, and $r$ fail to satisfy the triangle rule.

The literature distinguishes between trivial and nontrivial zeros of the 6-$j$ symbol.  Trivial zeros are zero because $N$ is zero, which occurs whenever one or more of the triples $\{(j_1 j_2 j_3), (\ell_1 \ell_2 j_3), (j_1 \ell_2 \ell_3), (\ell_1 j_2 \ell_3)\}$ fails to satisfy the triangle rule.  In other words, trivial zeros correspond to matrix elements that vanish as prescribed by known selection rules.  Nontrivial zeros are zero despite $N$ being nonzero, because the terms in the summation sum to zero.  This is possible because the terms in the summation have alternating sign.  Nontrivial zeros are classified by the number of terms in the summation, with the {\it weight} defined as the number of terms minus one.

The 6-$j$ symbol zero that is relevant to the case at hand, $\left\{ \begin{array}{ccc} 2 & 2 & 2 \\ \tfrac{3}{2} & \tfrac{3}{2} & \tfrac{3}{2} \end{array} \right\}=0$, is a nontrivial zero, since $N$ is nonzero, and is of weight 1, since $\alpha_{\rm max}=6$ and $\beta_{\rm min}=7$.  It turns out to be the canonical nontrivial zero, because it is the simplest one.  (They are enumerated in the literature and ours appears first in such lists \cite{bie84b} as it involves the smallest input parameters.)  Plugging in $j_1=j_2=j_3=2$ and $\ell_1=\ell_2=\ell_3=\tfrac{3}{2}$, Eqs.~(\ref{eq:6jgeneral}) through (\ref{eq:6jDelta}) yield
\begin{equation}
\left\{ \begin{array}{ccc} 2 & 2 & 2 \\ \tfrac{3}{2} & \tfrac{3}{2} & \tfrac{3}{2} \end{array} \right\} = \frac{2^5}{\sqrt{14}(6!)^2}\left[7! - \frac{8!}{(2!)^3}\right] = 0
\label{eq:6jours}
\end{equation}
Thus, the happenstance responsible for the vanishing of the 2-2 term boils down to the fact that $2^3 = 8$.

\section{Quadrupole-Quadrupole Energy Spectrum}
\label{sec:quadquad}

\subsection{Quadrupole-Quadrupole Interaction}
\label{ssec:interaction}
As noted in Sec.~\ref{ssec:intmodel}, the large-$R$ acceptor-acceptor interaction reduces to the electrostatic interaction between two quadrupoles separated in space by vector ${\bf R}$.  Placing quadrupole 1 at the origin, it is straightforward to show \cite{jac75} that its electrostatic potential at position ${\bf R} = x_1 \hat{\bf x} + x_2 \hat{\bf y} + x_3 \hat{\bf z} = R \hat{\bf R}$ is, in Gaussian units,
\begin{equation}
V_1({\bf R}) = \frac{e}{2 \epsilon_0 R^5} \sum_{kl} Q_{kl}^1 x_k x_l
\label{eq:electricpotential}
\end{equation}
where $Q_{kl}^1$ are the components of the quadrupole tensor of quadrupole 1 and indices $k$ and $l$ run from 1 to 3.  It is also straightforward to show \cite{jac75} that the potential energy cost of placing quadrupole 2 at position ${\bf R}$ in the presence of an arbitrary electrostatic potential $V({\bf R})$ is
\begin{equation}
U_2({\bf R}) = \frac{e}{6} \sum_{ij} Q^2_{ij} \frac{\partial^2 V}{\partial x_i \partial x_j}
\label{eq:potentialenergy}
\end{equation}
where $Q_{ij}^2$ are the components of the quadrupole tensor of quadrupole 2 and indices $i$ and $j$ run from 1 to 3.  Plugging $V_1$ in for $V$ yields (in agreement with Ref.~\onlinecite{han85}) the interaction energy of two quadrupoles separated by vector ${\bf R}$:
\begin{eqnarray}
U_{12}({\bf R}) = \frac{e^2}{6 \epsilon_0 R^5} \bigg[ && \!\!\!\!\!\! \sum_{ij} Q_{ij}^1 Q_{ij}^2 - 10 \sum_{ijk} n_i Q_{ij}^1 Q_{jk}^2 n_k \nonumber \\
&& \!\!\!\!\!\! + \; \tfrac{35}{2} \sum_{ij} n_i Q_{ij}^1 n_j \sum_{kl} n_k Q_{kl}^2 n_l \bigg]
\label{eq:interactioncomponent}
\end{eqnarray}
where $n_i \equiv x_i/R$ and we have made use of the fact that the quadrupole tensors are symmetric and traceless.  Converting to the units we have used throughout the rest of this paper (energies in effective Rydbergs, lengths in effective Bohr radii), rewriting in matrix form, and noting that the above is precisely our interaction Hamiltonian, we obtain
\begin{eqnarray}
H_{\rm int} = \frac{1}{3R^5} \Big[ && \!\!\!\!\!\! \tensor{\bf Q}^1 : \tensor{\bf Q}^2 - 10 \hat{\bf R}^{\rm T} \tensor{\bf Q}^1 \tensor{\bf Q}^2 \hat{\bf R} \nonumber \\
&& \!\!\!\!\!\! + \; \tfrac{35}{2} \left(\hat{\bf R}^{\rm T} \tensor{\bf Q}^1 \hat{\bf R}\right)\left(\hat{\bf R}^{\rm T} \tensor{\bf Q}^2 \hat{\bf R}\right) \Big]
\label{eq:interactionvector}
\end{eqnarray}
where $\hat{\bf R}$ is the unit (column) vector pointing from quadrupole 1 to quadrupole 2, $\hat{\bf R}^{\rm T}$ is its transpose, and $\tensor{\bf Q}^1 : \tensor{\bf Q}^2 \equiv \sum\limits_{ij} Q^1_{ij} Q^2_{ij}$ is the Frobenius product of the two quadrupole tensors.

\subsection{Matrix Elements of the Interaction Hamiltonian}
\label{ssec:hamiltonian}
Evaluating the interaction Hamiltonian, $H_{\rm int}$, in our basis of sixteen product states, $|F_{z1} F_{z2}\rangle = |\Psi_{F_{z1}}\rangle |\Psi_{F_{z2}}\rangle$, yields a $16\times16$ Hamiltonian matrix.  Since all terms in $H_{\rm int}$ are proportional to the product of a component of $\tensor{\bf Q}^1$ times a component of $\tensor{\bf Q}^2$, and since each quadrupole tensor acts only on its own single-acceptor states (in the large-separation limit), evaluation of matrix elements in the product-state basis is straightforward.  Matrix element $\langle F'_{z1} F'_{z2} | H_{\rm int} | F_{z1} F_{z2} \rangle$ is obtained by substituting $\langle \Psi_{F'_{z1}} | \tensor{\bf Q}^1 |\Psi_{F_{z1}} \rangle$ and $\langle \Psi_{F'_{z2}} | \tensor{\bf Q}^2 |\Psi_{F_{z2}} \rangle$ for $\tensor{\bf Q}^1$ and $\tensor{\bf Q}^2$ in Eq.~(\ref{eq:interactionvector}).

\begin{figure}
\centerline{\resizebox{3.5in}{!}{\includegraphics{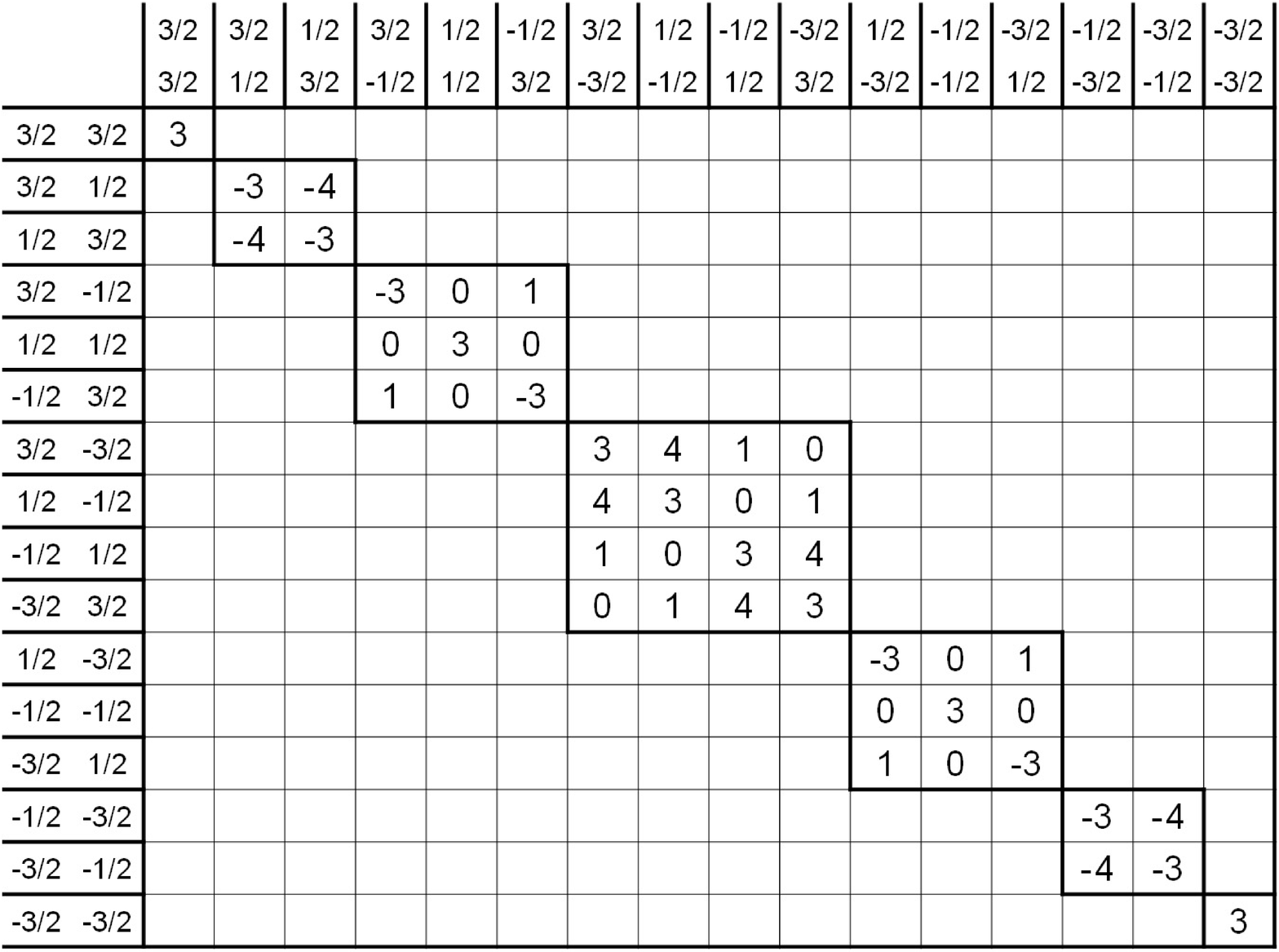}}}
\caption{Matrix elements of the quadrupole-quadrupole interaction Hamiltonian between acceptor-pair product states, $\langle F'_{z1} F'_{z2} | H_{\rm int} | F_{z1} F_{z2} \rangle$, are obtained by multiplying this $16\times16$ matrix by a common factor of $Q_0^2/R^5$.  All empty boxes contain zeros.  Defining $\hat{\bf z}$ along $\hat{\bf R}$ and arranging the product state basis in order of decreasing $F_z^{\rm tot}$ yields the block diagonal structure, with each block labeled by a distinct value of $F_z^{\rm tot}$: $3$, $2$, $1$, $0$, $-1$, $-2$, $-3$, from top left to bottom right.}
\label{fig:hamiltonianmatrix}
\end{figure}

The spherical symmetry of the Baldereschi-Lipari single-acceptor Hamiltonian [Eq.~(\ref{eq:hamiltonian})] means that we are free to define the quantization axis ($z$-axis) along any direction in coordinate space.  The simplest choice is to define it along the line joining the two acceptors.  Doing so sets $\hat{\bf R} = \hat{\bf z}$ in Eq.~(\ref{eq:interactionvector}), makes explicit the cylindrical symmetry of the acceptor-pair problem, and therefore requires the conservation of $F_z^{\rm tot} \equiv F_{z1} + F_{z2}$.  Thus, with this choice, product states of different $F_z^{\rm tot}$ cannot couple to each other, which sets all but 44 of the 256 matrix elements of $H_{\rm int}$ to zero.  Judicious ordering of the product-state basis yields a Hamiltonian matrix with the block diagonal form shown in Fig.~\ref{fig:hamiltonianmatrix}.  Additional symmetries (swapping up for down, swapping acceptor 1 for acceptor 2, and the hermiticity of the Hamiltonian) reduce the remaining 44 matrix elements to 13 unique ones that need to be calculated.  We calculate them by plugging the appropriate quadrupole tensors from Fig.~\ref{fig:tensorsmatrix} into Eq.~(\ref{eq:interactionvector}) and thereby obtain the Hamiltonian matrix, which is given by an overall factor of $Q_0^2/R^5$ times the matrix in Fig.~\ref{fig:hamiltonianmatrix}.  Thus,
\begin{equation}
\langle F'_{z1} F'_{z2} | H_{\rm int} | F_{z1} F_{z2} \rangle = \frac{Q_0^2}{R^5} \times [{\rm Fig.~\ref{fig:hamiltonianmatrix}}]
\label{eq:Hintmatrix}
\end{equation}
where $Q_0$ is the $\mu$-dependent quadrupole moment defined in Eq.~(\ref{eq:Q0}) that is equal to the absolute value of the expectation value of $Q_{zz}$ in any of the four single-acceptor ground states.  Due to the simple form of the matrix elements of the quadrupole tensor (see Sec.~\ref{sec:quadtensor}), all dependence on spin-orbit parameter (via $Q_0(\mu)$) and inter-acceptor separation $R$ resides in the prefactor above, which simply multiplies a matrix of numbers.

\subsection{Energy Spectrum}
\label{ssec:spectrum}
The seven blocks of the Hamiltonian matrix in Fig.~\ref{fig:hamiltonianmatrix} are each labeled by a different value of $F_z^{\rm tot}$: $3$, $2$, $1$, $0$, $-1$, $-2$, and $-3$, from top left to bottom right. Due to the up-down symmetry of the acceptor-pair, blocks labeled by $F_z^{\rm tot}$ and $-F_z^{\rm tot}$ are equivalent.  Thus, we can solve for the sixteen energy eigenvalues by diagonalizing the $1\times1$, $2\times2$, $3\times3$, and $4\times4$ submatrices corresponding to $F_z^{\rm tot}$ = $\pm3$, $\pm2$, $\pm1$, and $0$, noting that energies obtained from the first three submatrices are doubly degenerate.  Doing so, we obtain the following eight-level energy spectrum
\begin{equation}
E_{\rm int} = \frac{Q_0^2}{R^5} \times \left\{
\begin{array}{ccccccc}
8 &&&& \{0\} && (1)\\
6 &&&& \{0\} && (1)\\
3 &&&& \{-3,-1,1,3\} && (4)\\
1 &&&& \{-2,2\} && (2)\\
0 &&&& \{0\} && (1)\\
-2 &&&& \{-1,0,1\} && (3)\\
-4 &&&& \{-1,1\} && (2)\\
-7 &&&& \{-2,2\} && (2)
\end{array}
\right.
\label{eq:energyspectrum}
\end{equation}
where $F_z^{\rm tot}$ labels are provided within curly brackets and energy level degeneracies are listed in parentheses.  Note that while symmetry limits the number of distinct energy levels to ten, two additional, accidental degeneracies (the equivalence of a $0$-level and a $\pm1$-level at $E=-2Q_0^2/R^5$ and the equivalence of a $\pm1$-level and a $\pm3$-level at $E=3Q_0^2/R^5$) have reduced the number of distinct energy levels from ten down to eight.

\begin{figure}
\centerline{\resizebox{3.5in}{!}{\includegraphics{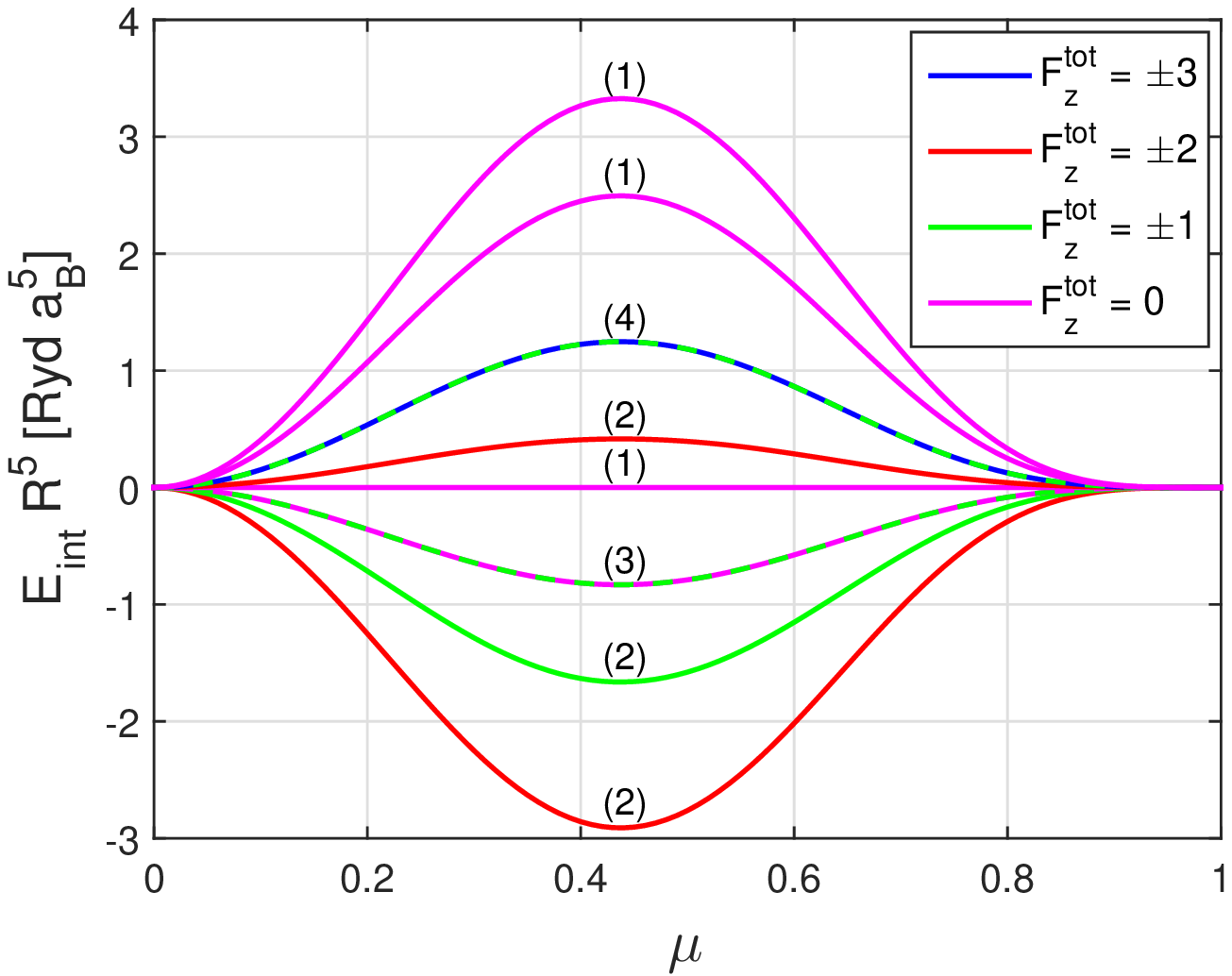}}}
\caption{Quadrupole-quadrupole interaction energy spectrum as a function of spin-orbit coupling parameter $\mu$.  We have plotted $E_{\rm int} R^5$ in order to remove the $1/R^5$ dependence on inter-acceptor separation $R$.  Line colors denote different $F_z^{\rm tot}$ quantum number labels.  Two-color dashed lines are labeled by both colors.  Level degeneracy is indicated in parentheses.  All $\mu$ dependence derives from a $Q_0^2$ prefactor, so there is no level crossing as a function of $\mu$.}
\label{fig:EvsMu}
\end{figure}

\begin{figure}
\centerline{\resizebox{3.5in}{!}{\includegraphics{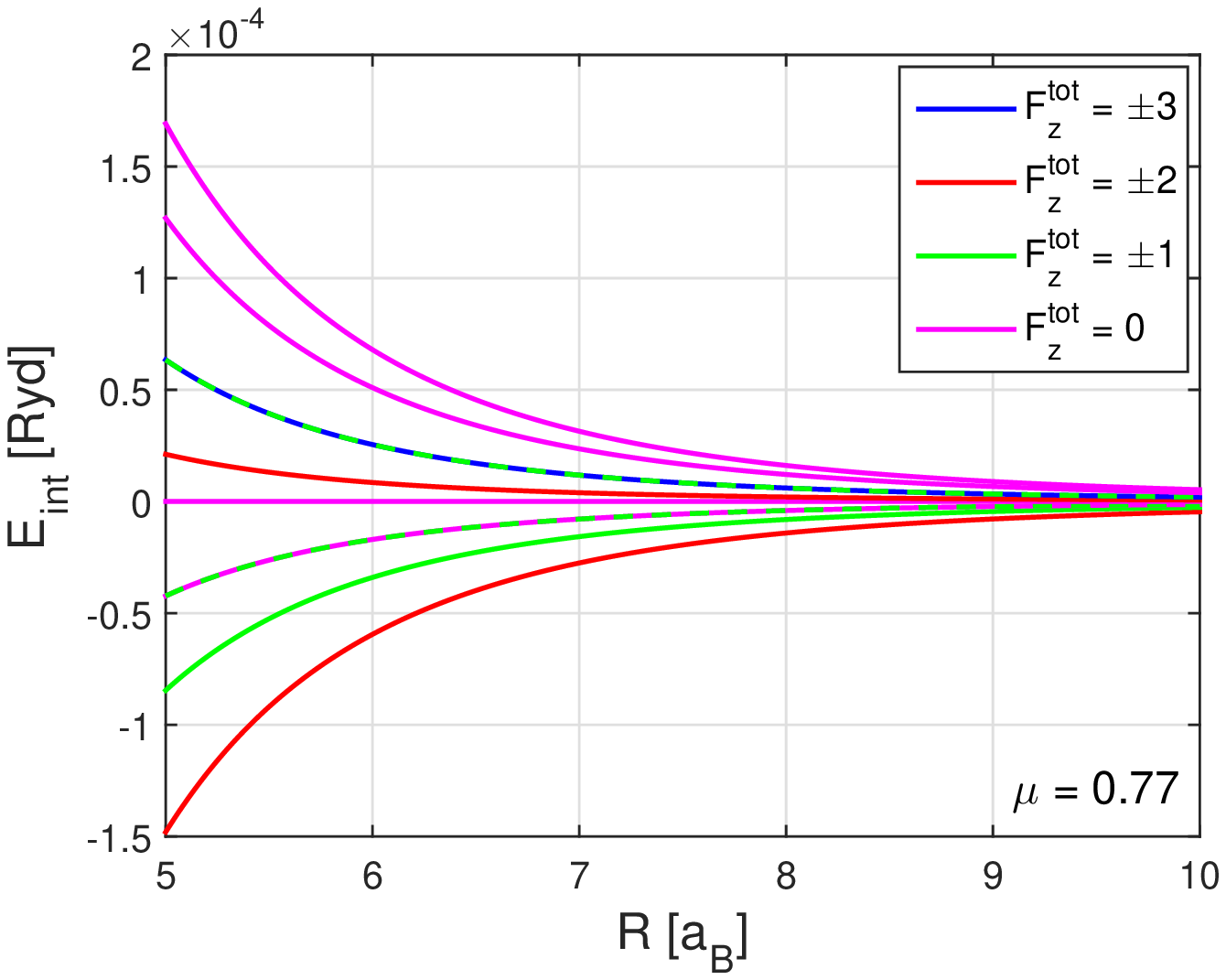}}}
\caption{Quadrupole-quadrupole interaction energy spectrum as a function of inter-acceptor separation $R$, for $\mu = 0.77$, which is the spin-orbit coupling parameter for GaAs.  Line colors denote different $F_z^{\rm tot}$ quantum number labels.  Two-color dashed lines are labeled by both colors.  All $R$ dependence derives from a $1/R^5$ prefactor, so there is no level crossing as a function of $R$.}
\label{fig:EvsRplot}
\end{figure}

This is clearly a far richer energy spectrum than the singlet-triplet spectrum of the hydrogen molecule, with eight levels instead of two.  And the ground state is two-fold degenerate, with $F_z^{\rm tot}=\pm2$, in contrast to the nondegenerate singlet ground state of the hydrogen molecule.  That said, the structure of this energy spectrum is surprisingly universal, with only the prefactor depending on the spin-orbit parameter $\mu$ or inter-acceptor separation $R$.  We plot the energy spectrum as a function of $\mu$ in Fig.~\ref{fig:EvsMu} and as a function of $R$ in Fig.~\ref{fig:EvsRplot}.  Since all $\mu$ and $R$ dependence enters through the prefactor, there is no level crossing as a function of these parameters. Due to the spherical symmetry of the Baldereschi-Lipari single acceptor Hamiltonian, this spectrum is also independent of $\hat{\bf R}$, the direction from one acceptor to the other.  If one sets $\hat{\bf R} \neq \hat{\bf z}$ in Eq.~(\ref{eq:interactionvector}), the Hamiltonian matrix is different and more complicated, but its eigenvalues are the same.

We can therefore characterize the quadrupole-quadrupole interaction by a single interaction parameter
\begin{equation}
J(R,\mu) \equiv \frac{Q_0(\mu)^2}{R^5}
\label{eq:J}
\end{equation}
with the seven excited-state energy levels separated from the ground state by $3J$, $5J$, $7J$, $8J$, $10J$, $13J$, and $15J$, respectively.  Since $Q_0$ is a function of $\mu$, $J$ is in units of effective Rydbergs, and $R$ is in units of effective Bohr radii, the interaction strength, measured via the splitting between the ground state and the first excited state, $3J$, is material dependent.  In Ge, for example, a splitting of 1~$\mu$eV is obtained for $R$=42~nm.  The same 1~$\mu$eV splitting is obtained in GaAs for $R$=24~nm, in Si for $R$=20~nm, in InP for $R$=19~nm, and in ZnTe for $R$=12~nm (based on parameter values from Table~I of Ref.~\onlinecite{bal73}).  (Note that our assumptions of large spin-orbit coupling and small cubic corrections are less valid for Si than the other materials.)  The dependence of the splitting on inter-acceptor separation, in each material, then follows from the $1/R^5$ factor in Eq.~(\ref{eq:J}).

\section{Conclusions}
\label{sec:conclusions}
We have developed a quadrupole-quadrupole model to describe the interaction between well-separated acceptors in doped semiconductors.  We modeled individual acceptors via the four-fold degenerate ground-state wave functions of the Baldereschi-Lipari spherical model \cite{bal73}.  Since such acceptors lack monopole or dipole moments but have nonzero quadrupole moments, the dominant interaction, for large inter-acceptor separation $R$, is the electric quadrupole-quadrupole interaction.  We calculated the matrix elements of the quadrupole tensor as a function of spin-orbit coupling parameter $\mu$.  Results were far simpler than expected, with all $\mu$-dependence entering via a single prefactor, $Q_0(\mu)$, that multiplies all tensor components for all matrix elements.  The form of this prefactor is further simplified by a mathematical happenstance, the nontrivial vanishing of a particular Wigner 6-$j$ symbol, $\left\{ \begin{array}{ccc} 2 & 2 & 2 \\ \tfrac{3}{2} & \tfrac{3}{2} & \tfrac{3}{2} \end{array} \right\} = 0$.  Deriving the quadrupole-quadrupole interaction Hamiltonian as a function of the two quadrupole tensors, we calculated its matrix elements between acceptor-pair product states and diagonalized to find the sixteen-state energy spectrum, with eigenstates labeled by $F_z^{\rm tot}$ quantum numbers.  Due to the simplicity of the quadrupole-tensor matrix elements, we were able to calculate this acceptor-pair energy spectrum in closed form.  It is an eight-level spectrum [see Eq.~(\ref{eq:energyspectrum})] controlled by a single interaction parameter, $J(R,\mu) = Q_0(\mu)^2 / R^5$, where the seven excited-state energy levels are separated from the ground state by $3J$, $5J$, $7J$, $8J$, $10J$, $13J$, and $15J$, respectively.  From low energy to high, the degeneracy of these eight levels is 2, 2, 3, 1, 2, 4, 1, and 1.  In contrast to the singlet ground state of the hydrogen molecule, the ground state here is two-fold degenerate, corresponding to $F_z^{\rm tot}=\pm2$.  Since all $R$ and $\mu$ dependence enters via $J$, there are no level crossings as a function of either parameter.  The acceptor-acceptor interaction is strongest for intermediate spin-orbit coupling, vanishing in both the $\mu\rightarrow0$ and $\mu\rightarrow1$ limits where the acceptor quadrupole moment vanishes.  It decays as a $1/R^5$ power law with increasing inter-acceptor separation and, within the spherical acceptor model, it is independent of the direction, $\hat{\bf R}$, from one acceptor to the other.

In the future, we plan to improve upon these results by including cubic corrections to the Baldereschi-Lipari spherical acceptor model [due to the previously neglected $\delta$ terms in Eq.~(\ref{eq:cubichamiltonian})].  Though such corrections are typically small, they are larger than typical in silicon and therefore of interest.  The effect of small cubic corrections on the single-acceptor eigenstates was studied perturbatively, by Baldereschi and Lipari, in Ref.~\onlinecite{bal74}.  While some single-acceptor excited-state degeneracies are split by cubic corrections, the single-acceptor ground states (on which we have based the present calculation) remain four-fold degenerate, shifting together in energy.  The single-acceptor ground-state wave functions, however, can become more complex in the presence of cubic corrections, with angular momentum states of all even $L$ contributing, not just the $L$=0 and $L$=2 states of Eq.~(\ref{eq:groundstate}).  These additional terms have the potential to modify the acceptor quadrupole moment and to alter the acceptor-pair energy spectrum by shifting energy levels and lifting the accidental degeneracies discussed in Sec.~\ref{ssec:spectrum}.  And we expect their inclusion to introduce some anisotropy to the acceptor-acceptor interaction, making the interaction dependent on the direction, $\hat{\bf R}$, from one acceptor to the other.  It will therefore be instructive to study such effects, both perturbatively in $\delta/\mu$, as should be sufficient for most semiconductors since $\delta/\mu \ll 1$, and nonperturbatively if possible, as may be necessary for Si where $\delta/\mu \approx 1/2$.

We also intend to apply our results to the development of a strong disorder renormalization group \cite{das80,bha81,bha82,fis94} (SDRG) scheme for studying the thermodynamic properties of acceptor-doped semiconductors \cite{roy86}.  The Bhatt-Lee SDRG technique \cite{bha81,bha82}, developed to study donor-doped semiconductors, assumes a two-level interaction spectrum where the singlet-triplet splitting decays exponentially with inter-donor separation.  We plan to generalize their technique to the eight-level spectrum derived herein, with $J$ decaying as a $1/R^5$ power law, in order to apply it to acceptor-doped semiconductors.

We are hopeful that the surprisingly simple, closed-form results derived herein will serve as a guide for experimenters who are working to control two-qubit interactions in quantum computing implementations based on acceptor spins.  The large inter-acceptor separation limit that we have considered should be directly relevant to the dilute dopant concentrations typical in such experiments.

\begin{acknowledgments}
The authors are grateful to B. Burrington for very helpful discussions regarding the matrix elements of spherical tensor operators and the nontrivial zeros of 6-$j$ symbols.  A.C.D. and G.Y.M. were supported by funds provided by Hofstra University, including a Faculty Research and Development Grant, a Presidential Research Award Program grant, and faculty startup funding.  R.N.B. acknowledges support from the Department of Energy, Office of Basic Energy Sciences, through grant DE-SC0002140.
\end{acknowledgments}

\appendix

\section{Wigner-Eckart Theorem}
\label{app:wigner}
The Wigner-Eckart theorem \cite{wig27,eck30,wig59,edm57,bie84a,sak17} states that for any spherical tensor operator $T_k^q$ of rank $k$
\begin{equation}
\langle \alpha^\prime j^\prime m^\prime | T_k^q | \alpha j m \rangle = \langle m q | j^\prime m^\prime \rangle_{jk} \langle \alpha^\prime j^\prime || T_k || \alpha j \rangle
\label{eq:wignereckart}
\end{equation}
where $|\alpha j m\rangle$ is an angular momentum eigenstate of azimuthal quantum number $j$, magnetic quantum number $m$, and additional quantum numbers collectively labeled $\alpha$, $\langle m q | j^\prime m^\prime \rangle_{jk}$ is the Clebsch-Gordan coefficient from the $j \times k$ Clebsch-Gordan table connecting the $|j m\rangle |k q\rangle$ uncoupled state to the $|j^\prime m^\prime\rangle$ coupled state, and $\langle \alpha^\prime j^\prime || T_k || \alpha j \rangle$ is known as the reduced matrix element, which is notably independent of $m$, $m^\prime$, and $q$.  Since the five $\ell$=2 spherical harmonics define a rank-2 spherical tensor, this applies directly to the case at hand.

For purely orbital angular momentum states, $|\alpha j m\rangle = |L L_z\rangle = Y_L^{L_z}(\theta,\phi)$, the matrix elements of spherical harmonics are the solid-angle integrals of the product of three spherical harmonics
\begin{eqnarray}
\langle L^\prime L_z^\prime | Y_\ell^m | L L_z \rangle &=& \int \left. Y_{L^\prime}^{L_z^\prime} \right.^* Y_\ell^m Y_L^{L_z} d\Omega \nonumber \\
&=& \langle m L_z | L^\prime L_z^\prime \rangle_{\ell L} \langle L^\prime || Y_\ell || L \rangle
\label{eq:orbitalwignereckart}
\end{eqnarray}
where the second equality is due to the Wigner-Eckart theorem.  Since such integrals have known solutions \cite{sak17}, the values of the reduced matrix elements in this special (purely orbital) case are known to be
\begin{equation}
\langle L^\prime || Y_\ell || L \rangle = \sqrt{\frac{(2\ell+1)(2L+1)}{4\pi(2L^\prime+1)}} \langle 0 0 | L^\prime 0 \rangle_{\ell L} \equiv C_{\ell L}^{L^\prime}
\label{eq:orbitalreducedmatrixelements}
\end{equation}
where $C_{\ell L}^{L^\prime}$ is a shorthand notation that we will use herein and we note that $C_{\ell L}^{L^\prime} = 0$ if $\ell + L + L^\prime$ is an odd integer.

For coupled (not purely orbital) angular momentum states, the Wigner-Eckart theorem still holds, though reduced matrix elements are not simply given by Eq.~(\ref{eq:orbitalreducedmatrixelements}).  For the $|L J F F_z\rangle$ states that appear in the Baldereschi-Lipari wave functions, $F$ is the azimuthal quantum number, $F_z$ is the magnetic quantum number, and $L$ and $J$ are the additional quantum numbers represented by $\alpha$ in Eq.~(\ref{eq:wignereckart}).  Thus, for these coupled states, the Wigner-Eckart theorem tells us that
\begin{equation}
\langle L^\prime J F F_z^\prime | Y_\ell^m | L J F F_z\rangle = \langle m F_z | F F_z^\prime \rangle_{\ell F} \langle L^\prime J F || Y_\ell || L J F \rangle
\label{eq:coupledwignereckart}
\end{equation}
where $\langle L^\prime J F || Y_\ell || L J F \rangle$ is the reduced matrix element which, very importantly, does not depend on $F_z$, $F_z^\prime$, or $m$.

In the matrix element calculations of Appendix~\ref{app:matrixelementcalculation}, we take advantage of both of the above applications of the Wigner-Eckart theorem.

\section{Matrix Element Calculation}
\label{app:matrixelementcalculation}
In this Appendix, we calculate the four coupled-state matrix elements of $\ell$=2 spherical harmonics that appear in Eq.~(\ref{eq:matrixelement}), referred to as the 0-0, 2-0, 0-2, and 2-2 terms. Using Clebsch-Gordan coefficients, we expand the coupled eigenstates $|L J F F_z\rangle$ in the uncoupled basis of product states $|L L_z\rangle |J J_z\rangle$
\begin{eqnarray}
|0 \tfrac{3}{2} \tfrac{3}{2} F_z\rangle &\equiv& |0 \; 0\rangle_L |\tfrac{3}{2} \; F_z \rangle_J
\label{eq:0ket}
\end{eqnarray}
\begin{eqnarray}
|2 \tfrac{3}{2} \tfrac{3}{2} F_z\rangle &\equiv&  \sum\limits_{L_z = -2}^2 \langle L_z \; F_z \text{-} L_z} | \tfrac{3}{2} \; F_z \rangle_{2 \tfrac{3}{2} \nonumber\\
&&\;\;\;\;\;\;\;\;\;\; \times \; |2 \; L_z\rangle_L \; |\tfrac{3}{2} \; F_z \text{-} L_z\rangle_J
\label{eq:2ket}
\end{eqnarray}
where the subscripts $L$ and $J$ clarify orbital angular momentum kets versus spin angular momentum kets and $\langle L_z \; F_z\text{-}L_z | \tfrac{3}{2} \; F_z\rangle_{2 \tfrac{3}{2}}$ is the Clebsch-Gordan coefficient found in the $2 \times \tfrac{3}{2}$ Clebsch-Gordan table that links the uncoupled state $|2 \; L_z\rangle |\tfrac{3}{2} \; F_z\text{-}L_z\rangle$ to the coupled state $|\tfrac{3}{2} \; F_z\rangle$.

The 0-0 term is easily shown to be zero.
\begin{eqnarray}
\langle 0 \tfrac{3}{2} \tfrac{3}{2} F'_z | Y_2^m | 0 \tfrac{3}{2} \tfrac{3}{2} F_z \rangle &=& \langle 0 \; 0  | Y_2^m | 0 \; 0 \rangle_L \langle \tfrac{3}{2} \; F'_z | \tfrac{3}{2} \; F_z \rangle_J \nonumber\\
&=& C_{20}^0 \; \langle m \; 0 | 0 \; 0 \rangle_{20} \; \delta_{F'_z, F_z} \nonumber\\
&=& 0
\label{eq:00term}
\end{eqnarray}
Here, the second equality is due to the Wigner-Eckart theorem, via Eq.~(\ref{eq:orbitalwignereckart}), as well as the orthonormality of the spin eigenstates.  The result is zero because the Clebsch-Gordan coefficient is zero, since the triple $(2,0,0)$ violates the triangle rule (one cannot add spin-2 to spin-0 and get spin-0).

Expanding the 2-0 term, we find that
\begin{eqnarray}
\lefteqn{\langle 2 \tfrac{3}{2} \tfrac{3}{2} F'_z | Y_2^m | 0 \tfrac{3}{2} \tfrac{3}{2} F_z \rangle} \nonumber \\
&=& \sum\limits_{L'_z = -2}^2 \langle L'_z \; F'_z\text{-}L'_z  | \tfrac{3}{2} \; F'_z \rangle_{2 \tfrac{3}{2}} \langle 2 \; L'_z | Y_2^m | 0 \; 0 \rangle_L \nonumber\\
&& \;\;\;\;\;\;\;\; \times \langle \tfrac{3}{2} \; F'_z\text{-}L'_z | \tfrac{3}{2} \; F_z \rangle_J \nonumber \\
&=& \sum\limits_{L'_z = -2}^2 \langle L'_z \; F'_z\text{-}L'_z | \tfrac{3}{2} \; F'_z \rangle_{2 \tfrac{3}{2}} \left[ C_{20}^2 \langle m \; 0 | 2 \; L'_z \rangle_{20} \right]
\delta_{L'_z, F'_z\text{-}F_z} \nonumber\\
&=& C_{20}^2 \langle F'_z\text{-}F_z \; F_z | \tfrac{3}{2} \; F'_z \rangle_{2 \tfrac{3}{2}} \langle m \; 0 | 2 \; F'_z\text{-}F_z\rangle_{20} \nonumber\\
&=& \tfrac{1}{\sqrt{4\pi}} \langle m \; F_z | \tfrac{3}{2} \; F'_z \rangle_{2 \tfrac{3}{2}}
\label{eq:20term}
\end{eqnarray}
where, once again, the second equality makes use of the Wigner-Eckart theorem [Eq.~(\ref{eq:orbitalwignereckart})] and the orthonormality of the spin eigenstates.  The fourth equality follows from using Eq.~(\ref{eq:orbitalreducedmatrixelements}) to compute $C_{20}^2 = 1/\sqrt{4\pi}$ and from noting that $\langle m \; 0 | 2 \; F'_z\text{-}F_z\rangle_{20} = \delta_{m,F'_z-F_z}$ and $\langle m \; F_z | \tfrac{3}{2} \; F'_z \rangle_{2 \tfrac{3}{2}} \propto \delta_{m,F'_z-F_z}$.  The matrix element is therefore only nonzero if $F'_z = F_z + m$.  Plugging in the Clebsch-Gordan coefficients from the $2 \times \tfrac{3}{2}$ table \cite{tab00} yields matrix element values for all $m$, $F_z$, and $F'_z$.  For each $m$, resulting values of $\langle 2 \tfrac{3}{2} \tfrac{3}{2} F'_z | Y_2^m | 0 \tfrac{3}{2} \tfrac{3}{2} F_z \rangle$ are given in row $F'_z = \{-3/2,-1/2,1/2,3/2\}$ and column $F_z = \{-3/2,-1/2,1/2,3/2\}$ of the following $4 \times 4$ matrices:
\begin{eqnarray}
m = -2 &\rightarrow& \tfrac{1}{5} \sqrt{\tfrac{5}{2\pi}}
\left[ \begin{array}{cccc}
	0 & 0 & 1 & 0\\
	0 & 0 & 0 & 1\\
	0 & 0 & 0 & 0\\
	0 & 0 & 0 & 0
	\end{array} \right] \nonumber \\
m = -1 &\rightarrow& \tfrac{1}{5} \sqrt{\tfrac{5}{2\pi}}
\left[ \begin{array}{cccc}
	0 & -1 & 0 & 0\\
	0 & 0 & 0 & 0\\
	0 & 0 & 0 & 1\\
	0 & 0 & 0 & 0
	\end{array} \right] \nonumber \\
m = 0 &\rightarrow& \tfrac{1}{5} \sqrt{\tfrac{5}{2\pi}}
\left[ \begin{array}{cccc}
	\tfrac{1}{\sqrt{2}} & 0 & 0 & 0\\
	0 & \tfrac{-1}{\sqrt{2}} & 0 & 0\\
	0 & 0 & \tfrac{-1}{\sqrt{2}} & 0\\
	0 & 0 & 0 & \tfrac{1}{\sqrt{2}}
	\end{array} \right] \nonumber \\
m = 1 &\rightarrow& \tfrac{1}{5} \sqrt{\tfrac{5}{2\pi}}
\left[ \begin{array}{cccc}
	0 & 0 & 0 & 0\\
	1 & 0 & 0 & 0\\
	0 & 0 & 0 & 0\\
	0 & 0 & -1 & 0
	\end{array} \right] \nonumber \\
m = 2 &\rightarrow& \tfrac{1}{5} \sqrt{\tfrac{5}{2\pi}}
\left[ \begin{array}{cccc}
	0 & 0 & 0 & 0\\
	0 & 0 & 0 & 0\\
	1 & 0 & 0 & 0\\
	0 & 1 & 0 & 0
	\end{array} \right]
\label{eq:matrices20term}
\end{eqnarray}

Expanding the 0-2 term, we see that
\begin{eqnarray}
\lefteqn{\langle 0 \tfrac{3}{2} \tfrac{3}{2} F'_z | Y_2^m | 2 \tfrac{3}{2} \tfrac{3}{2} F_z \rangle} \nonumber \\
&=& \sum\limits_{L_z = -2}^2 \langle L_z \; F_z\text{-}L_z  | \tfrac{3}{2} \; F_z \rangle_{2 \tfrac{3}{2}} \langle 0 \; 0 | Y_2^m | 2 \; L_z \rangle_L \nonumber\\
&& \;\;\;\;\;\;\;\; \times \langle \tfrac{3}{2} \; F'_z | \tfrac{3}{2} \; F_z\text{-}L_z \rangle_J \nonumber \\
&=& \sum\limits_{L_z = -2}^2 \langle L_z \; F_z\text{-}L_z | \tfrac{3}{2} \; F_z \rangle_{2 \tfrac{3}{2}} \left[ C_{22}^0 \langle m \; L_z | 0 \; 0 \rangle_{22} \right]
\delta_{L_z, F_z\text{-}F'_z} \nonumber\\
&=& C_{22}^0 \langle \text{-}m \; F'_z | \tfrac{3}{2} \; F_z \rangle_{2 \tfrac{3}{2}} \langle m \; \text{-}m | 0 \; 0\rangle_{22} \nonumber \\
&=& \tfrac{1}{\sqrt{4\pi}} \langle m \; F_z | \tfrac{3}{2} \; F'_z \rangle_{2 \tfrac{3}{2}}
\label{eq:02term}
\end{eqnarray}
where the final equality is obtained by taking advantage of the symmetries of the Clebsch-Gordan coefficients (see Eqs.~(3.5.14) to (3.5.16) of Ref.~\onlinecite{edm57}) to note that $C_{22}^0 = \sqrt{5} C_{20}^2$, $\langle m \; \text{-}m | 0 \; 0\rangle_{22} = \frac{(-1)^m}{\sqrt{5}} \langle m \; 0 | 2 \; m\rangle_{20}$, and $\langle \text{-}m \; F'_z | \tfrac{3}{2} \; F_z \rangle_{2 \tfrac{3}{2}} = (-1)^{-m} \langle m \; F_z | \tfrac{3}{2} \; F'_z \rangle_{2 \tfrac{3}{2}}$.  Comparing with Eq.~(\ref{eq:20term}), we see that the 0-2 term and the 2-0 term are equal:
\begin{equation}
\langle 0 \tfrac{3}{2} \tfrac{3}{2} F'_z | Y_2^m | 2 \tfrac{3}{2} \tfrac{3}{2} F_z \rangle = \langle 2 \tfrac{3}{2} \tfrac{3}{2} F'_z | Y_2^m | 0 \tfrac{3}{2} \tfrac{3}{2} F_z \rangle .
\label{eq:02equals20}
\end{equation}

Expanding the 2-2 term yields
\begin{eqnarray}
\lefteqn{\langle 2 \tfrac{3}{2} \tfrac{3}{2} F'_z | Y_2^m | 2 \tfrac{3}{2} \tfrac{3}{2} F_z \rangle} \nonumber \\
&=& \sum\limits_{L'_z \texttt{=-}2}^2 \sum\limits_{L_z \texttt{=-}2}^2 \langle L'_z\; F'_z \text{-} L'_z | \tfrac{3}{2} \; F'_z \rangle_{2\tfrac{3}{2}} \langle L_z \; F_z \text{-} L_z | \tfrac{3}{2} \; F_z \rangle_{2\tfrac{3}{2}} \nonumber\\
&& \;\;\;\;\;\;\;\; \times\langle 2 \; L'_z | \; Y_2^m \; | 2 \; L_z \rangle_L \langle \tfrac{3}{2} \; F'_{z} \text{-} L'_z | \tfrac{3}{2} \; F_z \text{-} L_z \rangle_J \nonumber\\
&=& \sum\limits_{L'_z \texttt{=-}2}^2 \sum\limits_{L_z \texttt{=-}2}^2 \langle L'_z\; F'_z \text{-} L'_z | \tfrac{3}{2} \; F'_z \rangle_{2\tfrac{3}{2}} \langle L_z \; F_z \text{-} L_z | \tfrac{3}{2} \; F_z \rangle_{2\tfrac{3}{2}} \nonumber\\
&& \;\;\;\;\;\;\;\; \times \left[ C_{22}^2 \langle m \; L_z | 2 \; L'_z \rangle_{22} \right] \delta_{L'_z, L_z \texttt{+} F'_z \text{-} F_z} \nonumber\\
&=&C _{22}^2 \sum\limits_{L_z \texttt{=-}2}^2 \langle L_z \texttt{+} m \; F_z \text{-} L_z | \tfrac{3}{2} \; F'_z \rangle_{2\tfrac{3}{2}} \langle L_z \; F_z \text{-} L_z | \tfrac{3}{2} \; F_z \rangle_{2\tfrac{3}{2}} \nonumber\\
&& \;\;\;\;\;\;\;\; \times \langle m \; L_z | 2 \; L_z \texttt{+} m \rangle_{22}
\label{eq:22term}
\end{eqnarray}
where, first, we expressed the coupled $|2 \tfrac{3}{2} \tfrac{3}{2} F_z\rangle$ states in the uncoupled product state basis via Eq.~(\ref{eq:2ket}).  Next, we used the Wigner-Eckart theorem, via Eq.~(\ref{eq:orbitalwignereckart}), to write the orbital matrix element $\langle 2 L'_z | Y_2^m | 2 L_z \rangle_L$ as the product of the reduced matrix element $C_{22}^2 = -\sqrt{\tfrac{5}{14\pi}}$ and the Clebsch-Gordan coefficient $\langle m L_z | 2 L'_z \rangle_{22}$.  Finally, we used the spin-state-orthonormality Kronecker delta to eliminate one of the sums, and we made use of the fact that the first Clebsch-Gordan coefficient vanishes unless $F'_z = F_z + m$.  What remains is the sum of products of three Clebsch-Gordan coefficients.  It is straightforward to calculate these matrix elements for any $m$, $F_z$, and $F'_z$ by looking up the coefficients in the $2 \times \tfrac{3}{2}$ and $2 \times 2$ tables and plugging in to the above.  But it turns out that, thanks to the Wigner-Eckart theorem, we need only calculate one of them.  Consider the $F'_z = F_z = \tfrac{1}{2}$ case.  We know from the start that this can only be nonzero for $m = \tfrac{1}{2} - \tfrac{1}{2} = 0$.  Plugging in from the Clebsch-Gordan tables \cite{tab00}, we find that this particular matrix element is zero.
\begin{eqnarray}
\lefteqn{\langle 2 \tfrac{3}{2} \tfrac{3}{2} \tfrac{1}{2} | Y_2^0 | 2 \tfrac{3}{2} \tfrac{3}{2} \tfrac{1}{2} \rangle} \nonumber \\
&=&C _{22}^2 \sum\limits_{L_z \texttt{=-}2}^2 \langle L_z \; \tfrac{1}{2} \text{-} L_z | \tfrac{3}{2} \; \tfrac{1}{2} \rangle_{2 \tfrac{3}{2}} \langle L_z \; \tfrac{1}{2} \text{-} L_z | \tfrac{3}{2} \; \tfrac{1}{2} \rangle_{2 \tfrac{3}{2}} \nonumber\\
&& \;\;\;\;\;\;\;\; \times \langle 0 \; L_z | 2 \; L_z \rangle_{22} \nonumber \\
&=& C_{22}^2 \left[ -\sqrt{\tfrac{2}{5}} \sqrt{\tfrac{2}{5}} \sqrt{\tfrac{1}{14}} - \sqrt{\tfrac{1}{5}} \sqrt{\tfrac{1}{5}} \sqrt{\tfrac{2}{7}} + \sqrt{\tfrac{2}{5}} \sqrt{\tfrac{2}{5}} \sqrt{\tfrac{2}{7}} \right] \nonumber\\
&=& \tfrac{-1}{7\sqrt{5\pi}} \left[ -1 - 1 + 2 \right] \nonumber\\
&=& 0
\label{eq:22termexample}
\end{eqnarray}
Recall from Appendix~\ref{app:wigner} that the Wigner-Eckart theorem can also be applied directly to matrix elements between coupled angular momentum states.  Thus, via Eq.~(\ref{eq:coupledwignereckart}), we can write
\begin{equation}
\langle 2 \tfrac{3}{2} \tfrac{3}{2} F'_z | Y_2^m | 2 \tfrac{3}{2} \tfrac{3}{2} F_z\rangle = \langle m F_z | \tfrac{3}{2} F'_z \rangle_{2 \tfrac{3}{2}}
\langle 2 \tfrac{3}{2} \tfrac{3}{2} || Y_2 || 2 \tfrac{3}{2} \tfrac{3}{2} \rangle
\label{eq:22coupledwignereckart}
\end{equation}
Since the last factor above, the reduced matrix element, is independent of $m$, $F_z$, and $F'_z$, it is the same for all cases and can be calculated from any case for which the Clebsch-Gordan coefficient is nonzero.  For the case we considered above, where $F'_z = F_z = \tfrac{1}{2}$ and $m=0$, the Clebsch-Gordan coefficient is $\langle 0 \tfrac{1}{2} | \tfrac{3}{2} \tfrac{1}{2} \rangle_{2 \tfrac{3}{2}} = -1/\sqrt{5} \neq 0$.  Thus, since we found that the matrix element is zero, the reduced matrix element must itself be zero.
\begin{equation}
\langle 2 \tfrac{3}{2} \tfrac{3}{2} || Y_2 || 2 \tfrac{3}{2} \tfrac{3}{2} \rangle =
\frac{\langle 2 \tfrac{3}{2} \tfrac{3}{2} \tfrac{1}{2} | Y_2^0 | 2 \tfrac{3}{2} \tfrac{3}{2} \tfrac{1}{2} \rangle}{\langle 0 \tfrac{1}{2} | \tfrac{3}{2} \tfrac{1}{2} \rangle_{2 \tfrac{3}{2}}}
= \frac{0}{\tfrac{-1}{\sqrt{5}}} = 0
\label{eq:22reducedmatrixelement}
\end{equation}
Plugging into Eq.~(\ref{eq:22coupledwignereckart}), we see that
\begin{equation}
\langle 2 \tfrac{3}{2} \tfrac{3}{2} F'_z | Y_2^m | 2 \tfrac{3}{2} \tfrac{3}{2} F_z\rangle = 0
\label{eq:22zero}
\end{equation}
for all $m$, $F_z$, and $F'_z$.  This is a surprising and remarkable result, which we make use of in Sec.~\ref{ssec:matrixelements} and discuss further in Sec.~\ref{ssec:vanish}.

\end{document}